\documentclass[aps,prl,floatfix,twocolumn]{revtex4}

\usepackage{epsfig}

\newcommand{\be}{\begin{equation}}
\newcommand{\ee}{\end{equation}}
\newcommand{\ba}{\begin{eqnarray}}
\newcommand{\ea}{\end{eqnarray}}
\newcommand{\nmax}{n_{\max}}
\begin{document}

%\twocolumn[\hsize\textwidth\columnwidth\hsize\csname @twocolumnfalse\endcsname
\title{Macroscopic magnetization jumps due to independent magnons\\
 in frustrated quantum spin lattices}

\author{J. Schulenburg$^1$, A. Honecker$^2$,
         J. Schnack$^3$, J. Richter$^1$
       and H.-J. Schmidt$^3$ }
\affiliation{
 $^1$ Institut f\"ur Theoretische Physik, Universit\"at Magdeburg,
      P.O. Box 4120, D-39016 Magdeburg, Germany
}
\affiliation{
%\address{
 $^2$ Institut f\"ur Theoretische Physik, TU Braunschweig,
      Mendelssohnstr. 3, D-38106 Braunschweig, Germany
}
\affiliation{
%\address{
 $^3$ Universit\"at Osnabr\"uck, Fachbereich Physik,
      Barbarastr. 7, D-49069 Osnabr\"uck, Germany
}
\date{August 29, 2001; revised February 25, 2002}

\begin{abstract}
For a class of frustrated spin lattices including the kagom\'e lattice
we construct exact eigenstates consisting of several independent,
localized one-magnon states and argue that they are ground states
for high magnetic fields. If the maximal number of local magnons scales
with the number of spins in the system, which is the case for the
kagom\'e lattice, the effect persists in the thermodynamic limit and
gives rise to a macroscopic jump in the zero-temperature magnetization
curve just below the saturation field. The effect decreases with increasing
spin quantum number and vanishes in the classical limit. Thus it is a
true macroscopic quantum effect.

\end{abstract}

%]

\maketitle

In frustrated quantum spin lattices the competition of quantum
and frustration effects promises rich physics. A reliable description
of such systems often constitutes a challenge for theory.
A famous example is the kagom\'e lattice antiferromagnet. In
spite of extensive studies during the last decade its ground
state properties are not fully understood yet. Classically it
has infinite continuous degeneracies. In the quantum case
($s$=1/2), the system is likely to be a spin liquid with a gap
for magnetic excitations and a huge number of singlet states
below the first triplet state (see
\cite{lecheminant97,waldtmann98,mila98} and references therein).

In this Letter we will focus on the zero-temperature
magnetic behavior of highly
frustrated lattices, in particular for high magnetic fields. One
aspect is given by the observation of nontrivial magnetic
plateaus in frustrated two dimensional (2D) quantum antiferromagnets like
SrCu$_2$(BO$_3$) \cite{kageyama99,onizuka00}, which has stimulated
theoretical interest (see e.g.\ \cite{MoTo}). Also the kagom\'e
lattice has a plateau at one third ($m=1/3$) of the saturation
magnetization \cite{hida01,CGHP}. 
Since this plateau can be found also in the Ising model and in the   
classical Heisenberg model with additional thermal fluctuations 
\cite{ZHP} it can be considered to be of classical origin.
However, the structure of the
ground state in the classical model is highly non-trivial at
$m=1/3$ \cite{mzh01} and has not been clarified yet for the
quantum model.

Another aspect is given by unusual jumps seen in magnetization
curves. Such jumps can arise for different reasons. One
possibility is a first-order transition between different
ground states like the spin flop transition in classical magnets
or in strongly anisotropic quantum chains \cite{gerhardt98}.
Here we discuss another possibility, namely a macroscopically
large degeneracy in the exact ground states of the full quantum
system for a certain value of the applied field.
We argue that this is a general phenomenon in highly frustrated
systems. This is remarkable in so far as one can {\it exactly}
write down ground states at a finite density of magnons in a
strongly correlated system which is neither integrable, nor has
any apparent non-trivial conservation laws. Such jumps
represent a genuine macroscopic quantum effect which is also of
possible experimental relevance since it occurs in many
well-known models like the kagom\'e lattice. This jump occurs
just below saturation and should be observable in magnetization
experiments on the corresponding compounds if the coupling
constants are small enough to make the saturating field
accessible.

We consider $N$ quantum spins of ``length'' $s$ described by the Hamiltonian
\begin{equation}
\label{Ham1}
 \hat{H} = \sum_{\langle ij \rangle} J_{ij}
\left\{\Delta \hat{S}_i^z \hat{S}_j^z + {1 \over 2} \left(
\hat{S}_i^{+} \hat{S}_j^{-} + \hat{S}_i^{-} \hat{S}_j^{+} \right) \right\}
       - h \hat{S}^z,
\end{equation}
where the sum runs over neighboring sites $\langle ij \rangle$,
$\hat{S}^z$ is the $z$-component of the total spin
$\hat{S}^z = \sum_i\hat{S}^z_i$,
$h$ the magnetic field, $\Delta$ the $XXZ$-anisotropy and
$J_{ij}$ are the exchange constants.

If the magnetic field $h$ is sufficient large, the ground state of
(\ref{Ham1}) becomes the magnon vacuum state $\vert 0\rangle
=\vert\uparrow\uparrow\uparrow\ldots\uparrow\uparrow\uparrow\rangle$
where
all spins assume their maximal $S_i^z$-quantum number.
The lowest excitations for the case of a high magnetic field are one-magnon
states $|1\rangle$. They are represented by states where the
$S^z$-quantum number is lowered by one and
can be written as
\be
    {}\vert 1\rangle=\frac{1}{c}\sum_i^{N}a_i\hat{S}_i^-\vert 0\rangle,
 \label{eq1magn}
\ee
where $c$ is chosen such that $\langle 1 \vert 1 \rangle = 1$.
For a lattice with $n_s$ spins per unit cell, there are
$n_s$ possible magnon bands $w_i(\vec{k})$ with $i=0,\ldots,(n_s-1)$.
For certain combinations of $J_{ij}$
of the Hamiltonian (\ref{Ham1})
the lowest magnon dispersion $w_0(\vec{k})$ becomes flat ({\it i.e.}\
independent of $k$) in some directions or in the whole $k$-space.

If the one-magnon dispersion is independent of one of the components
of $\vec{k}$, one can use Fourier transformation along this direction
to localize the one-magnon excitation along this direction in space.
If the dispersion is completely flat, the magnon can be localized
in a finite region of the lattice. This localized excitation can
have $N/n_s$ different positions. Now it is clear that one can
construct further local excitations. There will be no interaction
with the other excitations as long as they are sufficiently well
separated in space and therefore each excitation will have the
same energy. In this manner, one obtains $n$-magnon excitations
for $n \le \nmax$ whose energy is exactly $n$ times the one-magnon
energy. Due to the absence of attractive interaction, it is plausible
that these excitations are also the lowest $n$-magnon excitations. A proof
of this statement for $s=1/2$, arbitrary $\Delta \ge 0$ and all $J_{ij}$
equal will be given elsewhere \cite{schnack01} -- below we will report
numerical evidence for the occurrence of this effect for several models.
The essence of the above argumentation is that the ground state energies in
the 1-, 2-, \dots, $\nmax$-magnon spaces depend linearly on the
number of magnons, {\it i.e.}\ on the total magnetic quantum number. Hence the
total $S^z$ in the ground state goes directly from $Ns-\nmax$ to the
saturation value $Ns$ when increasing the magnetic field. In terms of
the magnetization curve $m(h)=S^z(h)/(Ns)$, this implies that there is
a jump $\delta m = \nmax/(Ns)$. If one band is completely flat, the
system can support a macroscopic amount of independent magnons $\nmax \sim N$
and one obtains a macroscopic jump just below saturation.

To be more precise, denote the region of localization of the magnon
state by $L$. Then the coefficients $a_l$ in (\ref{eq1magn}) are
different from zero only for sites $l\in L$.
The local one-magnon state is completely decoupled from the
rest of the lattice $R$ and the eigenstate $|\Psi\rangle$
can be written as a product
${}|\Psi\rangle=|\Psi_L\rangle|\Psi_R\rangle$
of a local part $L$ and the rest $R$.
$|\Psi_L\rangle$ is the local magnon state
and $|\Psi_R\rangle$ is the vacuum state.
The coefficients $a_i$ vanish for $i\in R$ in the one-magnon state
(\ref{eq1magn}),
{\it i.e.}\  $a_i \
 \neq 0 \;\: \forall i\in L$ and $a_i  = 0 \;\; \forall i\in R$.

The necessary and sufficient
condition for decoupling of the local state from the rest $R$ is
\be
  \label{cond1}
  \sum_{l\in L} a_l J_{lk} = 0 \qquad \forall \quad k\in R \, .
\ee
The Hamiltonian (\ref{Ham1}) can be divided into three parts
\ba  \hat{H} &=& \hat{H}_{L} + \hat{H}_{L-R} + \hat{H}_{R}
\ea
with $L$ being the part of the lattice where one magnon is
localized and $R$ the rest. The first term $\hat{H}_L$ is the local
part of the Hamiltonian with $J_{ij}=J_{l_1l_2}$ and $l_1,l_2 \in L$,
 whereof $|\Psi_L\rangle$ is the lowest
eigenstate. The second term $\hat{H}_{L-R}$ is the coupling of the
local part to the rest of the lattice with
$J_{ij}=J_{lk}$ and $l \in L$, $k\in R$. $J_{lk}$ must satisfy
condition (\ref{cond1}). The rest of the Hamiltonian which is
not connected with the local part is $\hat{H}_R$ with
$J_{ij}=J_{k_1k_2}$ and $k_1,k_2\in R$. $\hat{H}_{L-R}$ together
with condition (\ref{cond1}) creates the frustration,
therefore it seems that the magnetization jump described here is restricted to
highly frustrated lattices. The simplest realizations of such a
Hamiltonian are rings connected only by triangles.

%===================    figure   =================================
\begin{figure}[ht!]
\centerline{\epsfig{file=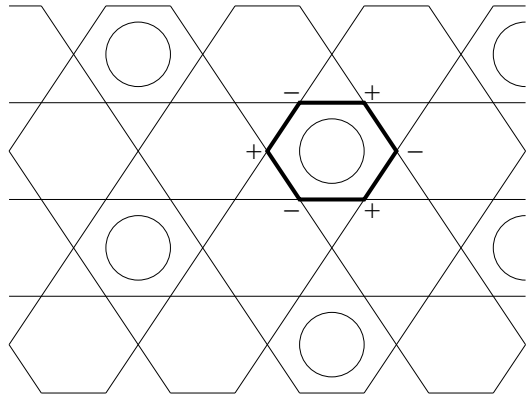,width=0.8\columnwidth}}
\vspace*{1mm}
\caption[]{Kagom\'e spin array, which hosts at least $N/9$ independent
magnons in a $\sqrt{3} \times \sqrt{3}$ structure if it consists of
$N$ spins. The circles mark hexagons where the independent magnons can be
localized. The structure of one localized magnon is indicated by
the signs around the bold hexagon which correspond to coefficient $a_l=\pm 1$
with which a spin-flip contributes at each site $l$.
Together with the surrounding triangles the local state satisfies equation
(\protect{\ref{cond1}}).}
\label{figfkago}
\end{figure}

In the kagom\'e lattice
\cite{lecheminant97,hida01}
as a typical example of flat one-magnon dispersion
$w_0(\vec{k}) = h-2 s J (1 + 2 \Delta)$
the magnon can be localized around a hexagon (see Fig.\ \ref{figfkago}).
Choosing the coefficients $a_l=(-1)^l$ with $l$ numbering the sites around a
hexagon ensures that $|\Psi_L\rangle$ is an exact eigenstate for the hexagon
(as illustrated by the bold one in Fig.\ \ref{figfkago}).
The triangles around the hexagon fulfill condition (\ref{cond1}) and
therefore
the magnon on the hexagon is decoupled from the rest of the lattice,
showing that this state is also an eigenstate of the whole kagom\'e lattice.
Further magnons can be put on the
lattice without disturbing existing excitations.
This can be repeated until every third hexagon is excited as shown in Fig.\
\ref{figfkago}. As a consequence we have a macroscopic magnetization jump
$\delta m$ having its
maximal value $\delta m = 2/9$ for the extreme quantum case $s=1/2$.

%===================    figure   =================================
\begin{figure}[ht!]
\centerline{\epsfig{file=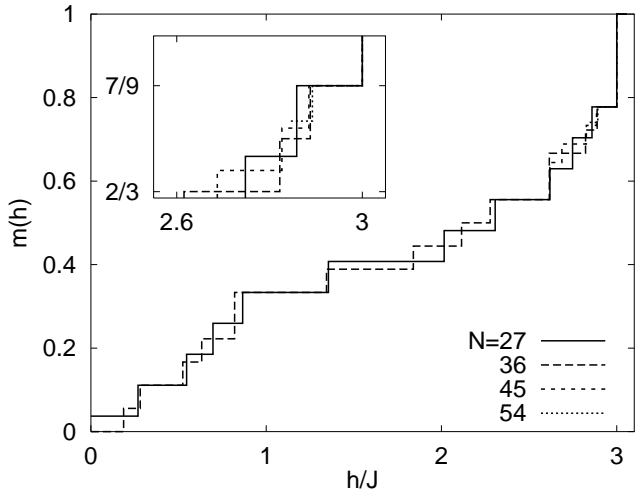,width=1.0\columnwidth}}
\vspace*{1mm}
\caption[]{Magnetization curves of an $s=1/2$ Heisenberg antiferromagnet on
a finite kagom\'e lattice for $N=27$, $36$, $45$, $54$ and $\Delta=1$.
For $N=45$ and $N=54$ the curve starts just below the magnetization jump.
It can be clearly seen that the jump to full magnetization has {\it no}
finite size dependence.
}
\label{figkagomag}
\end{figure}

A proof that these states are the lowest eigenstates in the
corresponding sectors is given in \cite{schnack01} for $s=1/2$,
arbitrary $\Delta \ge 0$ and all $J_{ij}$ equal. Here we present
numerical evidence for this statement. In Fig.\
\ref{figkagomag} exact diagonalization results are shown for
finite systems with periodic boundary conditions for the $s=1/2$
Heisenberg antiferromagnet on the kagom\'e lattice.  Only
lattices with $N$ being a multiple of 9 (three unit cells) are
presented, which fit to the $m=7/9$ state corresponding to Fig.\
\ref{figfkago}. The jump to saturation can be easily seen in this figure.
We have also computed curves on smaller clusters. They agree with those
presented in \cite{hida01,CGHP}. If the boundary conditions
do not fit to the $\sqrt{3}\times\sqrt{3}$ state or if the cluster is too
small, the jump may show finite-size effects.

We emphasize that the existence of this jump is quite independent of several
details of the system. First, one can construct $N/9$ independent magnons
for any $s$, leading to a jump of height $\delta m=1/(9s)$. Indeed, the
results for the $s=1$ kagom\'e lattice presented in \cite{hida01} show a
jump of the expected height for the given cluster sizes.
Second, introduction of an $XXZ$-anisotropy $\Delta \ne 1$ does not affect
the crucial properties of the one-magnon dispersion and therefore one
expects the magnitude of the degeneracy and the associated jump to be
independent of $\Delta$. In the exact diagonalization results
for $s=1/2$ \cite{CGHP} jumps of identical size are indeed observed
for $\Delta = 0$, $1$ and $2.5$. Third, the argumentation remains also
unchanged if one generalizes to different coupling constants in the
triangles pointing up and down (see Fig.\ \ref{figfkago}) \cite{mila98}.
The degeneracy is (partially) lifted only if coupling constants are changed
such that they become different around one triangle. The jump therefore
seems to be very stable not only in the kagom\'e lattice but also in the
other models to be discussed next where similar arguments can be applied.

Another 2D example for completely flat one-magnon dispersion
$w_0(\vec{k}) = h - 2 s J (1 + 3 \Delta)$ is the checkerboard
lattice \cite{canals01}, a 2D variant of the pyrochlore lattice.
In this case, localized magnon excitations
live around a square without diagonal interactions, again with
coefficients $a_l = (-1)^l$. The magnetization jump is $\delta m =1/(8s)$.
We have verified the predicted degeneracy and
associated macroscopic jump numerically for the checkerboard lattice
with $s=1/2$ and $\Delta = 1$.

Completely flat bands can also be found in dimensions different
from two. For example, the generalized pyrochlore lattice in
three dimensions with
two different coupling constants $J$, $J'$ (see e.g.\ \cite{canals98})
gives rise to the high frustration necessary
for a decoupling of local magnon excitations. The lowest
two out of the four
magnon bands are indeed degenerate and completely flat:
$w_0(\vec{k}) = w_1(\vec{k}) = h -s (J + J') (1 + 3 \Delta)$.
We expect a macroscopic jump of $\delta m\ge 1/(12s)$ for all $J,J'\ge 0$.

%===================    figure   =================================
\begin{figure}[ht!]
\centerline{\epsfig{file=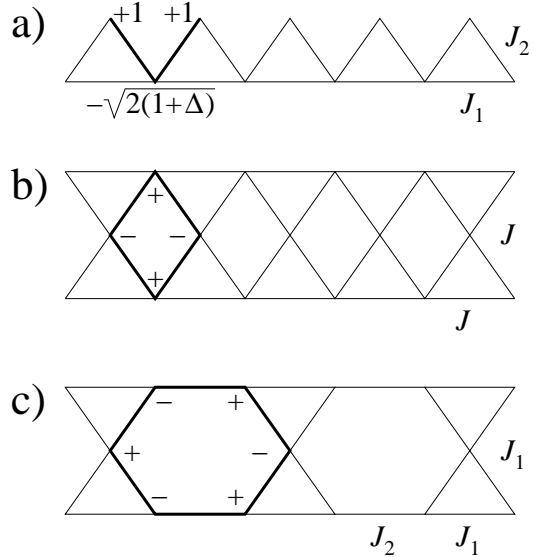,width=0.8\columnwidth}}
\vspace*{1mm}
\caption[]{One dimensional quantum antiferromagnets
with a magnetization jump to the full moment $m=1$:
 a) The sawtooth chain \protect{\cite{Naka95,sen95}},
 b) the kagom\'e like chain of \protect{\cite{wksme}} and
 c) the kagom\'e like chain of \protect{\cite{ahllt}}.
The position of a local magnon is marked by thick lines
with coefficients $a_l$ attached.
}
\label{fig1d}
\end{figure}

\begin{figure}[ht!]
\centerline{\epsfig{file=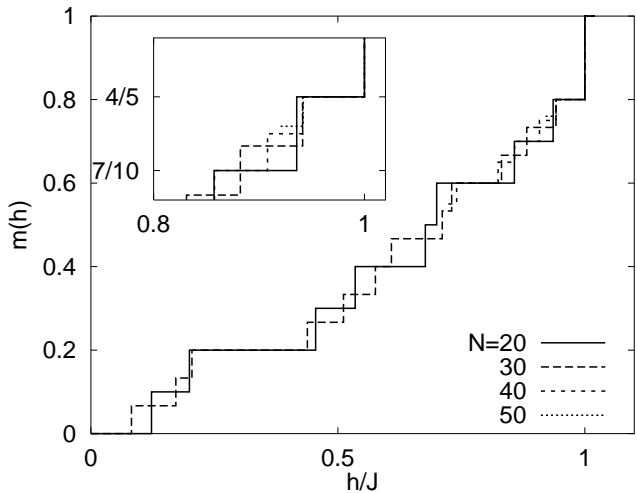,width=1.0\columnwidth}}
\vspace*{1mm}
\caption[]{Magnetization curves of the $s=1/2$ $XY$-kagom\'e
chain ($\Delta=0$) of Fig.\ \ref{fig1d}c) with $J_1 = J_2 = J$
for $N=20$, $30$, $40$, $50$. Inset: Region just below the jump to saturation.
}
\label{figkag1xy}
\end{figure}

Even in one dimension, one can find systems with a flat dispersion:
Some examples are shown in Fig.\ \ref{fig1d} together with the
structure of the localized magnon excitations.
For the generalized sawtooth chain \cite{Naka95,sen95} of Fig.\ \ref{fig1d}a),
the lowest magnon branch is completely flat for
$J_2 = \sqrt{2 (1 + \Delta)} J_1$ and all $s$.
Note that this example satisfies (\ref{cond1}) with more complicated
coefficients $a_l$ than encountered previously.
The lowest magnon branch for the one-dimensional kagom\'e
variant \cite{wksme} shown in Fig.\ \ref{fig1d}b)
is also completely flat $w_0(\vec{k}) = h -2 s J (1 + 2 \Delta)$.
Fig.\ \ref{fig1d}c) shows another variant of a kagom\'e chain
\cite{ahllt}. Here, the state indicated by the bold hexagon is an
eigenstate for $J_2 = (2 \Delta + 1) J_1 / (\Delta + 1)$
with $w_0(\vec{k}) = h -2 s J_1 (1 + 2 \Delta)$.
We have checked numerically that for $s=1/2$ and
$\Delta = 1$, $0$ a jump of size $\delta m = 1/2$, $1/3$ or
$1/5$ exists in case a), b) or c), respectively.
As an example, Fig.\ \ref{figkag1xy} shows the $m(h)$ curve of
the model of Fig.\ \ref{fig1d}c) with $s=1/2$, $\Delta = 0$, and
$J_2 = J_1$ (to ensure a flat dispersion). The jump of height
$\delta m = 1/5$ can be seen clearly (compare also the inset).
Furthermore, one can see several plateaus in the magnetization
curve. This suggests that the same conditions which give
rise to the jump also favor the formation of magnetization
plateaus. In particular, 
our numerical data always show a plateau preceding the jump.
However, this is
beyond the scope of the present Letter and needs further
investigations.

%===================    figure   =================================

So far, we have discussed cases with completely flat dispersion.
However, in more than one dimension it is also possible that the
lowest magnon branch has a flat dispersion only in some, but not
all directions. This has been noticed previously for the
$J_1-J_2$ model on the square lattice antiferromagnet with $J_2
= J_1/2$ \cite{honecker00}.
The generalized checkerboard lattice \cite{canals01} is another
2D model which has flat directions if the couplings $J'$ along
the diagonals are larger than those for the square lattice $J$.
In these cases, magnons remain delocalized along one dimension and thus
the localization region $L$ is a one-dimensional object.
For the $J_1-J_2$ model, $L$ extends along one of the axes of the square lattice
while for the checkerboard model it extends along one of the diagonals.
In both cases, $a_l = (-1)^l$ satisfies the condition (\ref{cond1}) and
yields localized magnon excitations. However, now the number of possible
magnon excitations is bounded by the linear extent of the lattice and
therefore their density tends to zero in the thermodynamic limit.
Consequently, the jump is only observable in finite systems
(this has been checked for both models with $s=1/2$ and $\Delta = 1$)
and vanishes in the thermodynamic limit. Arguments similar to those
used in \cite{yang97} then indicate that all derivatives $d^i h(m)/d m^i$
with $i \ge 1$ vanish at the point of full magnetization $m=1$. This
is in contrast to \cite{yang97} where $d^2 h(m)/d m^2$ was assumed
to be non-zero and thus a square root was inferred in $m(h)$ for
the $J_1-J_2$ model with $J_2 = J_1/2$.
Since the transition appears to become continuous for $N \to \infty$,
a non-analytic functional dependence in $h(m)$ like $h(m) \sim \exp(-1/(1-m))$
appears more plausible.
In any case, also in such a situation the
magnetization curve will be exceptionally steep close to saturation,
a fact which should also be detectable in high field experiments on
materials realizing such models.

In summary, we have shown
for certain frustrated spin lattices
that for magnetic fields near saturation
non-interacting magnons can condensate into a
single-particle ground-state leading to a macroscopic jump in the
magnetization curve.
This is a true quantum effect which vanishes if the spins become classical
($s\rightarrow\infty$).

Since this effect is generic in highly frustrated magnets, we
are confident that a realization of some model discussed in
the present Letter can be found with sufficient small $s$ and
$J$ to make the saturating field accessible.
The dynamics of the lattice may be relevant in such a compound,
but one will have to see whether this strengthens or weakens the
anomaly predicted at the saturating field. Our analysis of
deformations of the coupling constants at least indicates that the
low-temperature magnetization curve will show an unusually steep
rise even if the geometry deviates from the ideal structure.
It is interesting to note that a similar jump also appears
in certain molecular magnets \cite{schnack01} where the magnetic
ions form an icosidodecahedron (like \{Mo$_{72}$Fe$_{30}$\}), a
cuboctahedron or similar frustrated structures.

{\it   Acknowledgement:}
This work was partly supported by the DFG (project Ri615/10-1).

%{\it Note added in proof:}
% Recently, one of the present authors generalized
% the proof that the states constructed in this Letter are ground states
% to general $s$ \cite{schmidt02}.

%%%%%%%%%%%%%%%%%%%%%%%%%%%%%%%%%%%%%%%%%%%%%%%%%%%%%%%%%%%%%%%%%%%%%

\end{document}